# An Efficient Algorithm Based on Wavelet Transform to Reduce Powerline Noise From Electrocardiograms

Juan Ródenas[1], Manuel García[1], José J. Rieta[2], Raúl Alcaraz[1]

[1] Research Group in Electronic, Biomedical and Telecomm. Eng., Univ. of Castilla-La Mancha, Spain
[2] BioMIT.org, Electronic Engineering Department, Universitat Politecnica de Valencia, Spain

**Abstract**

*Nowadays, the electrocardiogram (ECG) is still the most widely used signal for the diagnosis of cardiac pathologies. However, this recording is often disturbed by the powerline interference (PLI), its removal being mandatory to avoid misdiagnosis. Although a broad variety of methods have been proposed for that purpose, often they substantially alter the original signal morphology or are computationally expensive. Hence, the present work introduces a simple and efficient algorithm to suppress the PLI from the ECG. Briefly, the input signal is decomposed into four Wavelet levels and the resulting coefficients are thresholded to remove the PLI estimated from the TQ intervals. The denoised ECG signal is then reconstructed by computing the inverse Wavelet transform. The method has been validated making use of fifty 10-min length clean ECG segments obtained from the MIT–BIH Normal Sinus Rhythm database, which were contaminated with a sinusoidal signal of 50 Hz and variable harmonic content. Comparing the original and denoised ECG signals through a signed correlation index, improvements between 10–72% have been observed with respect to common adaptive notch filtering, implemented for comparison. These results suggest that the proposed method is featured by an enhanced trade-off between noise reduction and signal morphology preservation*

## 1. Introduction

The surface electrocardiogram (ECG) has proven to be extremely useful in the diagnosis of many cardiac disorders, thus resulting essential in daily clinical practice [1,2]. This signal reflects the heart electrical activity through potential differences captured by electrodes placed in standardized positions on the patient's thorax. However, the potentials obtained in this way are extremely weak, usually presenting few millivolts. Thus, they can be substantially disturbed by different kinds of nuisance signals, including baseline wandering, electromyographic interferences or instrumentation noise from electronic devices. In this context, the powerline interference (PLI) is a major common perturbation in the ECG [3]. This nuisance signal is caused by capacitive coupling between the patient's body and the surrounding wirings from the mains or many others associated to power supplies of active equipments [4].

Although the nominal mains frequency is well-established in each country (50 or 60 Hz), it often presents fluctuations both in frequency and amplitude [5]. Hence, the PLI can be considered as a non-stationary nuisance signal, its successful removal being a challenging task [3].

Another aspect also hampering PLI reduction is the fact that this interference falls within the bandwidth of interest for the ECG recording, i.e. within the frequency range from 0.05 to 150 Hz [1]. Both issues have been broadly considered to explain the poor performance exhibited by the most common algorithms recently proposed for PLI reduction from the ECG. In fact, typical fixed-bandwidth notch filtering presents the major limitation of requiring a relatively wide stop-band to deal with frequency deviations in the PLI, thus notably disturbing the original ECG morphology [6]. To palliate this problem, several adaptive approaches with ability to track time-varying fluctuations in the PLI have also been proposed [7–9]. Although these methods have reported a better performance than common notch filtering, they still introduce a significantly large alteration in the ECG morphology. Indeed, sudden voltage transitions associated with QRS complexes often interfere with the parameter estimation [9].

As an alternative to these methods, denoising based on Wavelet transform (WT) has been recently applied to the ECG signal [10–14]. Thus, given a suitable wavelet function, it has been proven that orthogonal decomposition of a noisy ECG recording can separate its main profile from overlapped white noise and other sinusoidal interferences [13, 14]. This ECG transformation has been traditionally developed through a well-known multiresolution algorithm, where the input signal is progressively filtered and decimated [13]. However, this procedure does not preserve translation invariance in the resulting wavelet coeffi-cients, thus removing high-frequency information from the input signal and, then, modifying the native morphology of the QRS complexes [15, 16]. To minimize this aspect and achieve a good trade-off between PLI reduction and original ECG morphology preservation, a novel denoising algorithm based on stationary WT (SWT) is proposed in the present work. Note that this transformation is a shifted invariant version of the discrete WT, where the input signal is never sub-sampled and the filters are upsampled at each level of decomposition [15, 16].

## 2. Methods

### 2.1. Study population

To validate the proposed algorithm, fifty 10-min length and noise-free ECG segments were extracted from the MIT–BIH Normal Sinus Rhythm database, which is freely available at PhysioNet [17]. This dataset consists of 18 long-term ECG signals, which were acquired with a sampling rate of 128 Hz. Although two leads were available for all the recordings, only the one showing most common morphological patterns for the ECG waves was chosen.

The selected ECG segments were then resampled to 1000 Hz and contaminated with a synthesized PLI. To mimic this interference as realistic as possible, common fluctuations in the mains frequency were considered. According to the standard EN50160 [5], the power supply frequency is mainly set at 50 Hz with a maximum variation of ±1%. Moreover, this signal can also present harmonic components with a power lower than 2, 5, 1 and 6% for the first four multiples of 50 Hz, respectively [5]. Consequently, the PLI was simulated as a sinusoidal signal of 50 Hz and its first four harmonic components with random amplitude and frequency variations within the described limits. The resulting interference was finally used to obtain noisy ECG recordings with signal-to-interference (SIR) ratios of 15, 10, 5, 0, −5 and −10 dB.

### 2.2. SWT-based denoising algorithm

A block diagram summarizing the proposed denoising algorithm is displayed in Figure 1. As can be seen, the noisy ECG recording was firstly decomposed into four wavelet levels making use of a sixth-order Daubechies function. The resulting coefficients for the scales 1, 2, 3 and 4 were then thresholded to remove the PLI. As a final step, the denoised ECG signal was reconstructed by applying the inverse SWT to the shrunk wavelet coefficients.

In this methodology proper selection of shrinkage thresholds (i.e., $\lambda_1$, $\lambda_2$, $\lambda_3$ and $\lambda_4$) plays a key role to successfully cancel out the PLI and, simultaneously, to preserve the original ECG morphology. Indeed, the optimal cut-off value for each scale must act as an oracle to dis-cern between relevant and irrelevant wavelet coefficients. For this purpose, an adaptive threshold $\lambda_j$ for each scale was computed in this work by discarding high-amplitude information, which is mainly associated with QRS complexes [15, 16]. More precisely, the threshold $\lambda_j$ was obtained as the result of applying a moving median filter for a window of 200 ms to the absolute value of the wavelet coefficients for the scale $j$.

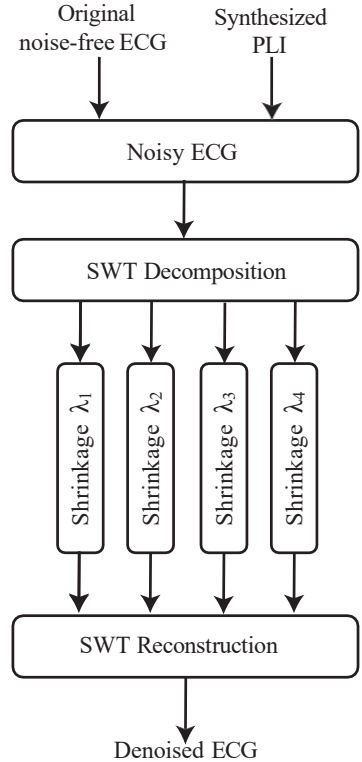

Figure 1. Block diagram for the proposed SWT-based denoising algorithm.

The way by which wavelet coefficients are thresholded also had a significant impact on the PLI suppression from the ECG signal. Thus, a combination of the well-known soft and hard thresholding functions was adopted to exploit the main characteristics of the ECG morphology [14]. In fact, whereas a hard thresholding was applied to the QRS complexes, a soft shrinkage was used for the remaining ECG intervals.

### 2.3. Performance assessment

As a reference for comparison, an adaptive notch filtering was also considered to reduce the PLI. This algorithm was implemented according to the indications found in the Costa et al.'s work [18]. Moreover, an adaptive signed correlation index (ASCI) was used to quantify noise reduction and morphology preservation in the resulting ECG signal for the two analyzed denoising algorithms. From a mathematical point of view, if the original clean ECG is named $x(n)$ and the denoised recording $\hat{x}(n)$, this metric was computed as

$$\text{ASCI}\big[x(n), \hat{x}(n)\big] = \frac{1}{L} \sum_{k=1}^{L} x(k) \otimes \hat{x}(k), \quad (1)$$

where $L$ is the number of samples both for $x(n)$ and $\hat{x}(n)$ and the operator $\otimes$ is defined as

$$x(n) \otimes \hat{x}(n) = \begin{cases} 1 & \text{if } |x(n) - \hat{x}(n)| \leq \xi, \\ -1 & \text{if } |x(n) - \hat{x}(n)| > \xi. \end{cases} \quad (2)$$

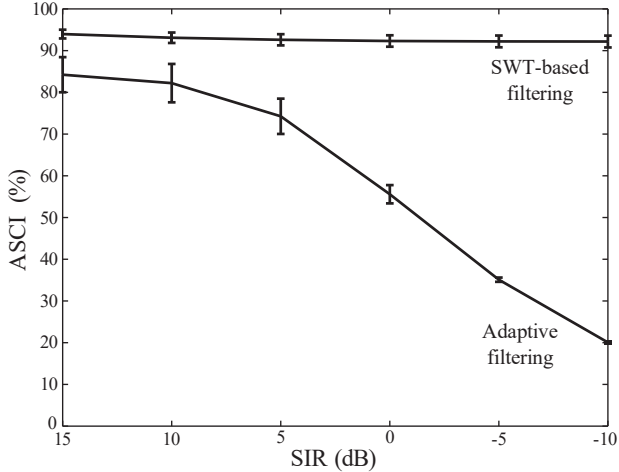

Figure 2. Mean and standard deviation values of ASCI obtained from the proposed SWT-based denoising algorithm and the reference adaptive notch filtering for different levels of SIR.

The threshold $\xi$ was experimentally set to 5% of the standard deviation of $x(n)$.

## 3. Results

Figure 2 summarizes average and standard deviation values of ASCI for the proposed SWT-based denoising algorithm and the reference adaptive notch filtering. As shown, whereas the proposed method presented a stable behavior regardless of the noise level, adaptive filtering was specially sensitive to SIR levels of ≤ 10 dB. Indeed, the SWT-based technique reported improvements of ASCI between 10–72% with respect to the reference filtering.

Figure 3 displays a denoising example with the two considered algorithms applied to a noisy ECG with SIR of 5 dB. Note how the resulting recording provided by the proposed method was clearly denoised preserving its native morphology, whereas a highly contaminated signal was still observed for the adaptive notch filtering.

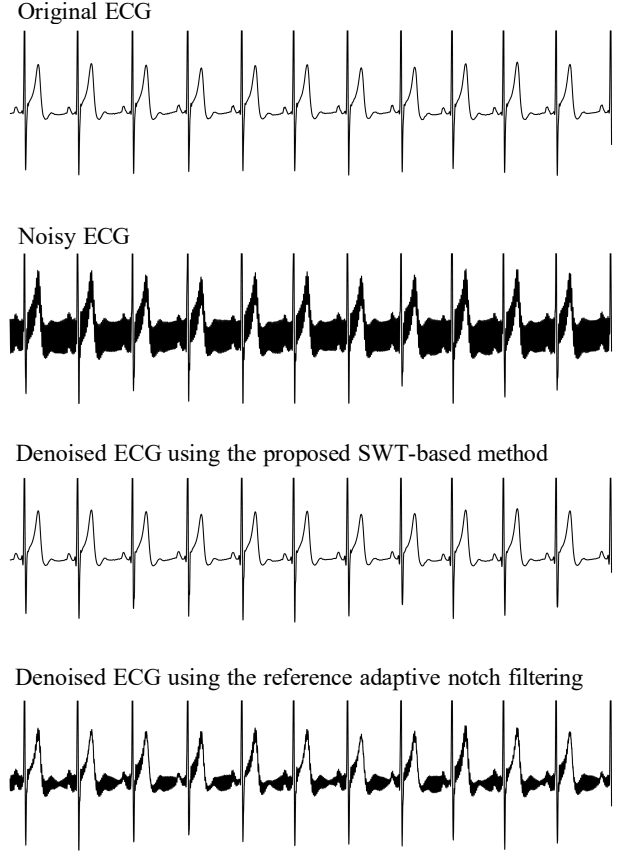

Figure 3. Typical example of the resulting ECG signals from the two analyzed denoising methods when a SIR of 5 dB was considered.

## 4. Discussion and conclusions

To a greater or lesser extent, the PLI is always present in the surface ECG recording and, therefore, a filtering stage able to reduce this perturbation is mandatory before further processing. For this purpose, adaptive notch filtering has been widely used as an alternative to common fixed-bandwidth filtering, due to its ability to track time-varying changes in the input signal [18]. However, although the results obtained in this study have shown an acceptable behavior for high values of SIR, the implemented adaptive filtering presented a poor performance for moderate and high levels of noise. Thus, whereas mean values of ASCI about 85% were seen for a SIR of 15 dB, they were around 20% for −10 dB (see Figure 2). This worsening performance with growing levels of noise has also been suggested by other authors as a major drawback of this denoising algorithm [7, 9].

Interestingly, this limitation has been largely overcome by the proposed SWT-based algorithm. As can be seen in Figure 2, morphological alterations caused by this de-noising method were negligible and remained mostly constant for every SIR. Indeed, regardless of the noise level, mean and standard deviation values of ASCI were always about 94% and 2%, respectively. As a consequence, the proposed method reached a good trade-off between PLI removal and preservation of the native ECG integrity. Other attractive feature of this method is its simplicity, so that it may be easily incorporated to commercial recording systems. Nonetheless, additional experiments to corroborate that aspect will be developed in the near future.

## Acknowledgements

Research supported by the grants DPI2017-83952-C3 MINECO/AEI/FEDER, UE and SBPLY/17/180501/000411 from Junta de Comunidades de Castilla-La Mancha.

Address for correspondence:

Juan Ro´denas Garc´ıa
E.S.I. Informa´tica, Campus Univ., 02071, Albacete, Spain
Phone: +34–967–599–200 Ext. 2556
Fax: +34–967–599–224
e–mail: juan.rodenas@uclm.es